\PassOptionsToPackage{svgnames}{xcolor}
\documentclass[12pt,reqno]{article}

\usepackage[numbers,sort&compress]{natbib}
\usepackage{amsthm, amsmath, amsfonts, amssymb, amscd, mathtools, youngtab, euscript, mathrsfs, verbatim, enumerate, multicol, multirow, bbding, color, babel, esint, geometry, tikz, tikz-cd, tikz-3dplot, array, enumitem, hyperref, thm-restate, thmtools, datetime, graphicx, tensor, braket, slashed, standalone, pgfplots, ytableau, subfigure, wrapfig, dsfont, setspace, wasysym, pifont, float, rotating, adjustbox, pict2e,array}
\usepackage{amsfonts}
\usepackage{longtable}
\usepackage{tabularx}
\usepackage{caption}
\usepackage{booktabs}
\usepackage{amsmath}
\usepackage{centernot}
\usepackage{etoolbox}
\usepackage{makecell}
\usepackage{blindtext}
\usepackage[all]{xy}
\usepackage[utf8]{inputenc}
\usepackage[normalem]{ulem}
\makeatletter
\numberwithin{equation}{section}
\numberwithin{figure}{section}
\usetikzlibrary{arrows, positioning, decorations.pathmorphing, decorations.pathreplacing, decorations.markings, matrix, patterns, snakes}
\hypersetup{colorlinks=true}
\hypersetup{linkcolor=black}
\hypersetup{citecolor=black}
\hypersetup{urlcolor=black}
\theoremstyle{plain}
\newtheorem*{thm*}{\protect\theoremname}
\theoremstyle{definition}
\newtheorem*{defn*}{\protect\definitionname}
\providecommand{\definitionname}{Definition}
\providecommand{\theoremname}{Theorem}
\makeatother
\tikzset{
  big arrow/.style={
    decoration={markings,mark=at position 1 with {\arrow[scale=1.5,#1]{>}}},
    postaction={decorate},
    shorten >=0.4pt},
  big arrow/.default=black}

\newcommand{\ta}{\theta}

    {\endtcolorbox}
\usepackage{mdframed}
\mdfdefinestyle{statementstyle}{
   frametitlerule = true,
   frametitlefont = {\normalfont\bfseries},
   frametitlebackgroundcolor = blue!10,
}
\mdtheorem[style=statementstyle]{statement}{String Universality}
\mdtheorem[style=statementstyle]{statement1}{Fundamental string}
\mdtheorem[style=statementstyle]{statement2}{Supergravity string = fundamental string}
\mdtheorem[style=statementstyle]{statement3}{Emergent string conjecture}
\mdtheorem[style=statementstyle]{statement4}{Sharpened Distance Conjecture~\cite{Etheredge:2022opl}}

\begin{document}
\title{
\vbox{
\baselineskip 14pt
\hfill \hbox{\normalsize KEK-TH-2507
}}  \vskip 1cm
Dualities from Swampland principles
}
\author{
Alek Bedroya$^{\diamond}\,$
and Yuta Hamada$^\star\,^\dagger$
\vspace{5mm}
    \\
    $^\diamond$\normalsize{\it Jefferson Physical Laboratory, Harvard University, Cambridge, MA 02138, USA,}
    \\
    $^\star$\normalsize{\it Theory Center, IPNS, High Energy Accelerator Research Organization (KEK),}\\
    \normalsize{\it\quad 1-1 Oho, Tsukuba, Ibaraki 305-0801, Japan,} \\
    $^\dagger$\normalsize{\it Graduate University for Advanced Studies (Sokendai), 1-1 Oho, Tsukuba, }
    \\
    \normalsize{\it\quad Ibaraki 305-0801, Japan}\vspace{1mm}  
}
\date{}
\maketitle

\begin{abstract}
    
    We initiate the program of bottom-up derivation of string theory dualities using Swampland principles. In particular, we clarify the relation between Swampland arguments and all the string theory dualities in $d\geq9$ dimensional supersymmetric theories. Our arguments center around the sharpened distance conjecture and rely on various other Swampland principles.
\end{abstract}

\tableofcontents

\section{Introduction}

One of the most remarkable features of string theory is the emergence of new weakly coupled descriptions at the infinite distance limits of the moduli space. This observation has been formulated more precisely in the Swampland distance conjecutre~\cite{Ooguri:2006in} (see, e.g. Refs.~\cite{Grimm:2018ohb,Lee:2018urn} for tests in string theory), which plays the central role in the Swampland program~\cite{Vafa:2005ui} (see also reviews~\cite{Palti:2019,vanBeest:2021lhn,Grana:2021zvf,Agmon:2022thq}). The emergent weakly coupled descriptions which appear at different corners of the moduli space are dual to each other. Therefore, the distance conjecture, at its core, is a general quantification of the universality of dualities in string theory. Indeed, the distance conjecture is sometimes referred to as the duality conjecture (e.g. Ref.~\cite{Montero:2022prj}). However, the precise logical relation between the distance conjecture and the various string dualities is not clear. The purpose of the paper is to fill this gap. We make this connection more precise by showing that refinements of distance conjecture can in fact explain dualities under several assumptions. 

We focus on the higher dimensional supergravity theories with $16$ or $32$ supercharges.
We use a variety of Swampland conjectures to find the relation between the conjectures and all the string dualities in $d\geq9$. 
Our arguments center around the sharpened distance conjecture~\cite{Etheredge:2022opl}, and rely on various other conjectures such as the BPS completeness hypothesis~\cite{Kim:2019vuc}.
This work shows that the Swampland principles, with some assumptions, capture the essence of the duality web, and perhaps must be viewed as fundamental.

Note that all the theories we consider are supergravities and not string theories. In other words, we will not assume that the UV completions of the mentioned supergravities are the known string theories.

The paper is organized as follows.
In Section~\ref{sec:conjectures}, the sharpened distance conjecture and its connection to the emergent string conjecture~\cite{Lee:2019wij,Lee:2019xtm}, our main tool in the paper, is reviewed.
In Section~\ref{s2}, the T-duality between the IIA and IIB theories is derived from the bottom-up perspective.
In Section~\ref{sec:duality_32supercharges}, other 11d/10d string dualities of theories with $32$ supercharges are discussed. 
Similarly, 10d/9d string dualities of theories with $16$ supercharges are argued in Section~\ref{sec:duality_16supercharges}. 
We clarify the relationship between the string dualities and the Swmapland conjectures.
Section~\ref{sec:conclusions} is devoted to the conclusions.
The technical details are provided in Appendix \ref{A1}, \ref{A2} and \ref{massive_vector}.
The result of Ref.~\cite{Bedroya2023} is reviewed in Appendix~\ref{IIp}.

\section{Sharpened distance conjecture}\label{sec:conjectures}

Let us start by reviewing the statements of the emergent string conjecture and the sharpened distance conjecture, as we will frequently use them throughout the paper. 

\begin{statement3*}
In any infinite distance limit of the field space, the lightest tower of states is either a KK tower coming from the higher dimensional field theory or excitations of a fundamental string \cite{Lee:2019wij,Lee:2019xtm}.
\end{statement3*}

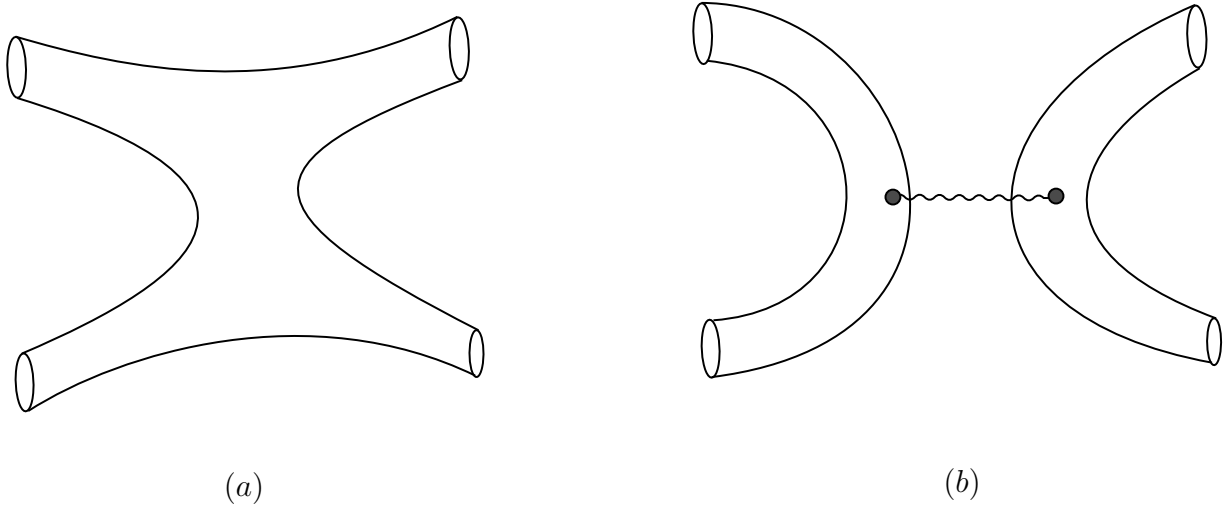
\begin{figure}[H]
    \centering

\tikzset{every picture/.style={line width=0.75pt}} 

\begin{tikzpicture}[x=0.75pt,y=0.75pt,yscale=-1,xscale=1]

\draw    (23.48,28) .. controls (77.99,44) and (157.36,61.91) .. (246.36,18) ;
\draw    (29.89,217) .. controls (77.19,186.62) and (172.75,159.05) .. (255.33,199.12) ;
\draw    (24.28,59) .. controls (170.99,103.83) and (117.28,149.47) .. (26.68,188) ;
\draw    (248.76,50) .. controls (135.72,92.52) and (138.33,115.63) .. (256.99,176.22) ;
\draw   (30.49,191.2) .. controls (32.59,196.29) and (33.2,205.48) .. (31.86,211.74) .. controls (30.52,218) and (27.72,218.95) .. (25.62,213.86) .. controls (23.52,208.78) and (22.91,199.58) .. (24.25,193.32) .. controls (25.59,187.07) and (28.39,186.12) .. (30.49,191.2) -- cycle ;
\draw   (26.62,31.55) .. controls (28.8,36.82) and (29.41,46.46) .. (27.99,53.08) .. controls (26.57,59.7) and (23.65,60.79) .. (21.47,55.52) .. controls (19.29,50.24) and (18.68,40.6) .. (20.1,33.98) .. controls (21.52,27.36) and (24.44,26.27) .. (26.62,31.55) -- cycle ;
\draw   (249.88,21.85) .. controls (252.14,27.32) and (252.8,37.18) .. (251.37,43.88) .. controls (249.93,50.58) and (246.93,51.57) .. (244.67,46.1) .. controls (242.41,40.63) and (241.74,30.77) .. (243.18,24.07) .. controls (244.62,17.38) and (247.62,16.38) .. (249.88,21.85) -- cycle ;
\draw   (258.03,178.42) .. controls (259.61,182.24) and (259.97,189.63) .. (258.84,194.92) .. controls (257.7,200.21) and (255.5,201.41) .. (253.92,197.58) .. controls (252.34,193.76) and (251.98,186.37) .. (253.12,181.08) .. controls (254.25,175.79) and (256.45,174.59) .. (258.03,178.42) -- cycle ;
\draw    (369.48,11) .. controls (477.5,11.5) and (536.5,179.5) .. (374.62,199.86) ;
\draw    (618.36,12) .. controls (492.5,66.5) and (494.5,168.5) .. (626.92,192.58) ;
\draw    (372.08,40.2) .. controls (464.5,49.5) and (466.5,164.5) .. (375.49,171.2) ;
\draw    (620.76,44) .. controls (553.5,82.5) and (533.5,134.5) .. (628.99,170.22) ;
\draw   (376.49,174.2) .. controls (378.59,179.29) and (379.2,188.48) .. (377.86,194.74) .. controls (376.52,201) and (373.72,201.95) .. (371.62,196.86) .. controls (369.52,191.78) and (368.91,182.58) .. (370.25,176.32) .. controls (371.59,170.07) and (374.39,169.12) .. (376.49,174.2) -- cycle ;
\draw   (372.62,14.55) .. controls (374.8,19.82) and (375.41,29.46) .. (373.99,36.08) .. controls (372.57,42.7) and (369.65,43.79) .. (367.47,38.52) .. controls (365.29,33.24) and (364.68,23.6) .. (366.1,16.98) .. controls (367.52,10.36) and (370.44,9.27) .. (372.62,14.55) -- cycle ;
\draw   (621.88,15.85) .. controls (624.14,21.32) and (624.8,31.18) .. (623.37,37.88) .. controls (621.93,44.58) and (618.93,45.57) .. (616.67,40.1) .. controls (614.41,34.63) and (613.74,24.77) .. (615.18,18.07) .. controls (616.62,11.38) and (619.62,10.38) .. (621.88,15.85) -- cycle ;
\draw   (630.03,172.42) .. controls (631.61,176.24) and (631.97,183.63) .. (630.84,188.92) .. controls (629.7,194.21) and (627.5,195.41) .. (625.92,191.58) .. controls (624.34,187.76) and (623.98,180.37) .. (625.12,175.08) .. controls (626.25,169.79) and (628.45,168.59) .. (630.03,172.42) -- cycle ;
\draw    (467,109) .. controls (468.67,107.35) and (470.34,107.36) .. (472,109.03) .. controls (473.66,110.7) and (475.33,110.71) .. (477,109.06) .. controls (478.68,107.41) and (480.35,107.42) .. (482,109.1) .. controls (483.66,110.77) and (485.33,110.78) .. (487,109.13) .. controls (488.67,107.48) and (490.34,107.49) .. (492,109.16) .. controls (493.66,110.83) and (495.33,110.84) .. (497,109.19) .. controls (498.67,107.54) and (500.34,107.55) .. (502,109.22) .. controls (503.66,110.89) and (505.33,110.9) .. (507,109.25) .. controls (508.68,107.6) and (510.35,107.61) .. (512,109.29) .. controls (513.66,110.96) and (515.33,110.97) .. (517,109.32) .. controls (518.67,107.67) and (520.34,107.68) .. (522,109.35) .. controls (523.66,111.02) and (525.33,111.03) .. (527,109.38) .. controls (528.67,107.73) and (530.34,107.74) .. (532,109.41) .. controls (533.65,111.09) and (535.32,111.1) .. (537,109.45) .. controls (538.67,107.8) and (540.34,107.81) .. (542,109.48) -- (545.5,109.5) -- (545.5,109.5) ;
\draw  [fill={rgb, 255:red, 74; green, 74; blue, 74 }  ,fill opacity=1 ] (462.25,109) .. controls (462.25,106.93) and (463.93,105.25) .. (466,105.25) .. controls (468.07,105.25) and (469.75,106.93) .. (469.75,109) .. controls (469.75,111.07) and (468.07,112.75) .. (466,112.75) .. controls (463.93,112.75) and (462.25,111.07) .. (462.25,109) -- cycle ;
\draw  [fill={rgb, 255:red, 74; green, 74; blue, 74 }  ,fill opacity=1 ] (544.5,108.5) .. controls (544.5,106.43) and (546.18,104.75) .. (548.25,104.75) .. controls (550.32,104.75) and (552,106.43) .. (552,108.5) .. controls (552,110.57) and (550.32,112.25) .. (548.25,112.25) .. controls (546.18,112.25) and (544.5,110.57) .. (544.5,108.5) -- cycle ;

\draw (127,246.4) node [anchor=north west][inner sep=0.75pt]    {$( a)$};
\draw (490,244.4) node [anchor=north west][inner sep=0.75pt]    {$( b)$};

\end{tikzpicture}
    \caption{(a) The worldsheet diagrams that contribute to the amplitude of string states. (b) For an ordinary defect that couples to spacetime fields, such diagrams need to be summed. However, for fundamental strings, including such diagrams would lead to overcounting of the amplitude since the spacetime fields are supposed to emerge from string perturbation.}
    \label{Diagrams}
\end{figure}

Note the word fundamental in describing the string. The criterion that the string is fundamental makes the conjecture very strong. What we mean by a fundamental string is a string that has graviton as a string state, and moreover, the scattering amplitudes of the string states are dominated by processes involving string worldsheets (see Fig.~\ref{Diagrams}). In other words, the scattering amplitudes of the string states are given by a string perturbation theory. 

The emergent string conjecture is a very strong refinement of the distance conjecture. However, we will be using a weaker refinement of the distance conjecture in our studies which is the sharpened distance conjecture \cite{Etheredge:2022opl}. The sharpened distance conjecture states that the value of the numerical coefficient $\lambda$ in the distance conjecture must be lower bounded by $1/\sqrt{d-2}$. Moreover, the authors in Ref.~\cite{Etheredge:2022opl} also made the observation that the inequality is only saturated when one of the string towers is the lightest tower. Our arguments will be based on the following conjecture.

\begin{statement4*}
The numerical coefficient $\lambda$ in the distance conjecture satisfies
\begin{align}
\lambda\geq\frac{1}{\sqrt{d-2}}=:\lambda_\text{min}\,.
\end{align}
The inequality is saturated if and only if the lightest tower is the string states.
\end{statement4*}

Note that in the above conjecture, it is not assumed that the corresponding string is necessarily fundamental. In fact, for this precise reason, the sharpened distance conjecture is a weaker conjecture than the emergent string conjecture. A derivation of the sharpened distance conjecture from the emergent string conjecture was given in Ref.~\cite{Agmon:2022thq}. One could formulate a weaker version of the emergent string conjecture without the assumption that the light string would be fundamental. The sharpened distance conjecture is stronger than the weak version of the emergent string conjecture and weaker than the emergent string conjecture.  
\begin{gather*}
   \text{Emergent string conjecture}\nonumber\\\Downarrow \nonumber\\
   \text{Sharpened distance conjecture}\nonumber\\
   \Downarrow\nonumber\\
   \text{Weak emergent string conjecture (without fundamentality of the string)}
\end{gather*}
In Ref.~\cite{Bedroya2023}, the sharpened distance conjecture was combined with the finiteness of the black hole entropy to show that the number of string limits is always countable. This is a powerful result that was used in Ref.~\cite{Bedroya2023} to argue for string universality in the strong coupling regime of 9d minimal supergravity. This result implies that with the exception of a measure zero subset of infinite distance limits, any other limit decompactifies. Here, we extend the use of this statement to find a bottom-up argument for dualities and infinite distance limits in the moduli space of the 9d theories. However, there might be countable infinite distance limits that our bottom-up argument misses, which are precisely the string limits or the interfaces between two different descriptions. 

\section{IIA/IIB T-duality}\label{s2}

Theories with $N\geq 16$ supercharges have at least one anti-symmetric two form in their gravity multiplet other than eleven dimension. Each 2-form $B_{\mu\nu}$ is a higher form gauge field with coupling $g$ which is set by the action,
\begin{align}\label{cdef}
    \mathcal{L}_B=\frac{1}{2g^2}dB\wedge\star dB.
\end{align}
A string that is electrically charged under $B_{\mu\nu}$ is called the supergravity string. The supergravity string is special in that it can be BPS. In all the known string theories, in the limit $g\rightarrow 0$, a BPS supergravity string becomes the fundamental string. Moreover, if we compactify the limit theory on a circle, it always has a T-dual in which the $g\rightarrow 0$ and $R\to0$.

In the following, we provide a bottom-up argument for this observation in theories with $32$ supercharges in $10$ dimension based on Swampland principles.
More concretely, we use the BPS completeness hypothesis~\cite{Kim:2019vuc}, the sharpened distance conjecture.

Assuming that the BPS supergravity string exists, the tension of the BPS supergravity string is (see Appendix \ref{A1} for derivation)\footnote{There are two $B_{\mu\nu}$ fields in the type IIB supergravity. Here we can choose any of them.}
\begin{align}
    T=
    M_{10}^2e^{\hat\phi/\sqrt{2}},
\end{align}
where $\hat{\phi}$ is the canonically normalized dilaton field, $M_{10}$ is the 10d Planck scale, and $g=e^{\sqrt{2}\,\hat\phi}$.

Let us define the string length $l_s$ such that,
\begin{align}
    l_s=T^{-\frac{1}{2}}=
    l_{10}e^{-\hat\phi/(2\sqrt{2})},
\end{align}
where $l_{10}=M_{10}^{-1}$ is the 10d Planck length.
We compactify the 10d theory on a circle with radius $R$ and consider a BPS version of a closed string, which is the winding string.

The mass of a winding string is protected by supersymmetry and is given by 
\begin{align}
    m_\text{winding}=TR=\frac{R}{l_s^2},
\end{align}
even for small values of $R$ where the 10d supergravity description breaks down. We can also find the degeneracies of the ground states of a winding string and their representations under the $SO(8)$ rotation group in 9d. This is because the ground states of the BPS string must furnish representations of the broken supersymmetries. In Appendix \ref{A2}, we show that based on the chirality of the 10d theory, the ground states of the winding string are given by

\begin{itemize}
    \item 
    Winding string in type\ IIA\ Supergravity: $(\mathbf{8}_v\oplus \mathbf{8}_s)\otimes(\mathbf{8}_v\oplus \mathbf{8}_s)$ with mass $m=\dfrac{R}{l_s^2}$.
    \item 
    Winding string in type\ IIB\ Supergravity: $(\mathbf{8}_v\oplus \mathbf{8}_c)\otimes(\mathbf{8}_v\oplus \mathbf{8}_s)$ with mass $m=\dfrac{R}{l_s^2}$, 
\end{itemize}
where $\mathbf{8}_v, \mathbf{8}_s$, and $\mathbf{8}_c$ are the vector, spinor, and conjugate spinor representations of Spin$(8)$, respectively.    

Recall that the massless spectrum of the IIA and IIB supergravity is given by $(\mathbf{8}_v\oplus \mathbf{8}_c)\otimes(\mathbf{8}_v\oplus \mathbf{8}_s)$ and $(\mathbf{8}_v\oplus \mathbf{8}_s)\otimes(\mathbf{8}_v\oplus \mathbf{8}_s)$.
After the compactification on $S^1$, these become the KK states charged under the KK $U(1)$.
On the other hand, the winding strings are charged under the $U(1)$ coming from the dimensional reduction of $B_{\mu\nu}$. 
Therefore, the first/second line is consistent with the first level of the KK tower of Type IIB/Type IIA supergravity on a circle with radius $l_s^2/R$. 
This is remarkable since the BPS completeness hypothesis alone seems to reproduce the well-known T-duality between type IIA and type IIB string theories. In particular, this almost shows that the limit $R\to0$ of IIA/IIB is the ten-dimensional IIB/IIA theory.

However, since we are not aware of the existence of the other non-BPS towers becoming light in the limit $R\to0$, there are still following possibilities.
\begin{enumerate}
\item The winding strings are the leading tower, corresponding to the decompactification limit to 10d.
\item The winding strings are the leading tower, corresponding to the decompactification limit to 11d.
\item The winding strings are the leading tower, corresponding to the tensionless string limit.
\item The winding strings are not the leading tower, and lighter non-BPS states are the leading tower, corresponding to the decompactification limit.
\item The winding strings are not the leading tower, and lighter non-BPS states are the leading tower, corresponding to the tensionless string limit.
\item The leading tower is neither KK nor string state.
\end{enumerate}
In the following, we show that the possibility $1$ is the only option using the sharpened distance conjectures.

First, it is easy to rule out the possibilities $2$ and $4$.
If the possibility $2$ is correct, given that we have maximal supersymmetry, the theory must decompactify to the 11d supergravity on $T^2$.
However, in this case, we have to find two BPS towers (the BPS tower other than the winding strings will be discussed in Section \ref{AM}). This does not occur at the generic point of $\theta$.

Similarly, the possibility $4$ is excluded. Since the decompactification limit is either 10d IIA/IIB supergravity on $S^1$ or 11d supergravity on $T^2$, the KK modes are always the BPS states, charged under $U(1)$ gauge symmetries. Therefore, it is impossible that the leading non-BPS tower becomes the KK tower.

Next, the sharpened distance conjecture excludes the possibility $6$ by definition.

Then, we use the sharpened distance conjecture to rule out the possibilities $3$ and $5$.
To this end, we clarify what we mean by $R\to0$ more precisely.
There are two moduli fields in 9d supergravity. One is the dilaton $\hat{\phi}$, and the other is the radion field parameterizing the $S^1$ radius.
The canonically normalized radion modulus in $d$ dimensions, $\hat\rho$, is given by\footnote{We treat $\hat{\rho}$ as a dimensionless quantity normalized by $l_9$.}
\begin{align}
    \frac{R}{l_9}=\exp\left[\sqrt{\frac{d-1}{d-2}}\,\hat{\rho}\right],
\end{align}
which becomes $R/l_9= e^{\sqrt{8/7}\,\hat{\rho}}$ for 9d.

We consider the infinite distance in $(\hat{\phi},\hat{\rho})$-plane.
\begin{align}
    &\hat{\phi}+i\hat{\rho}=:r e^{i\theta},
    &&r\to\infty,
    &&\theta:\text{fixed}.
\end{align}
The $R\to0$ limit corresponds to $\pi<\theta<2\pi$.
In this limit, $\hat{\phi}$ goes to $-\infty$ for $\pi<\theta<3\pi/2$, and goes to $+\infty$ for $3\pi/2<\theta<2\pi$. 
For $\theta=3\pi/2$, $\hat{\phi}$ remains constant.

As we do not want to change the effective 9d theory, we fix the value of $l_9$.
\begin{align}
    l_9=\left(l_s^8 e^{2\sqrt{2}\hat{\phi}}R^{-1}\right)^{1/7}
    =\left(l_s^8 e^{2\sqrt{2}\hat{\phi}-\sqrt{\frac{8}{7}}\hat{\rho}}/l_9\right)^{1/7}
    =\text{fixed},
\end{align}
where the 10d and 9d Planck lengths are related as $(l_9/l_{10})^8=l_9/R$.
This determines the behavior of $l_s$ in this limit:
\begin{align}
    &l_s\to l_9 \,e^{\frac{1}{2\sqrt{2}}\left(-\hat{\phi}+\frac{\hat{\rho}}{\sqrt{7}}\right)}
    =l_9 \,e^{\frac{1}{\sqrt{7}}\sin(\theta+\alpha)r},
    &&\tan\alpha=-\sqrt{7},
\end{align}
where $-\pi/2<\alpha<\pi/2$.

In this limit, the mass of the KK states and the BPS winding strings ($m_w=R/l_s^2$) become
\begin{align}
    &\frac{1}{R}=l_9^{-1}e^{-\sqrt{\frac{8}{7}}\hat{\rho}},
    \nonumber\\
    &\frac{R}{l_s^2}=l_9^{-1}e^{\left(\frac{\hat{\phi}}{\sqrt{2}}+\frac{3\hat{\rho}}{\sqrt{14}}\right)}
    =l_9^{-1}e^{\sqrt{\frac{8}{7}}\sin(\theta+\beta)r},
    &&\tan\beta=\frac{\sqrt{7}}{3},
\end{align}
where $-\pi/2<\beta<\pi/2$.
We observe that the numerical coefficient $\lambda_\text{winding}$ for the BPS winding string is given by
\begin{align}
    \lambda_\text{winding}=-\sqrt{\frac{8}{7}}\sin(\theta+\beta).
\label{eq:lambda}\end{align}

\begin{figure}[t]
    \begin{center}
        \includegraphics[width=80mm, height=60mm]{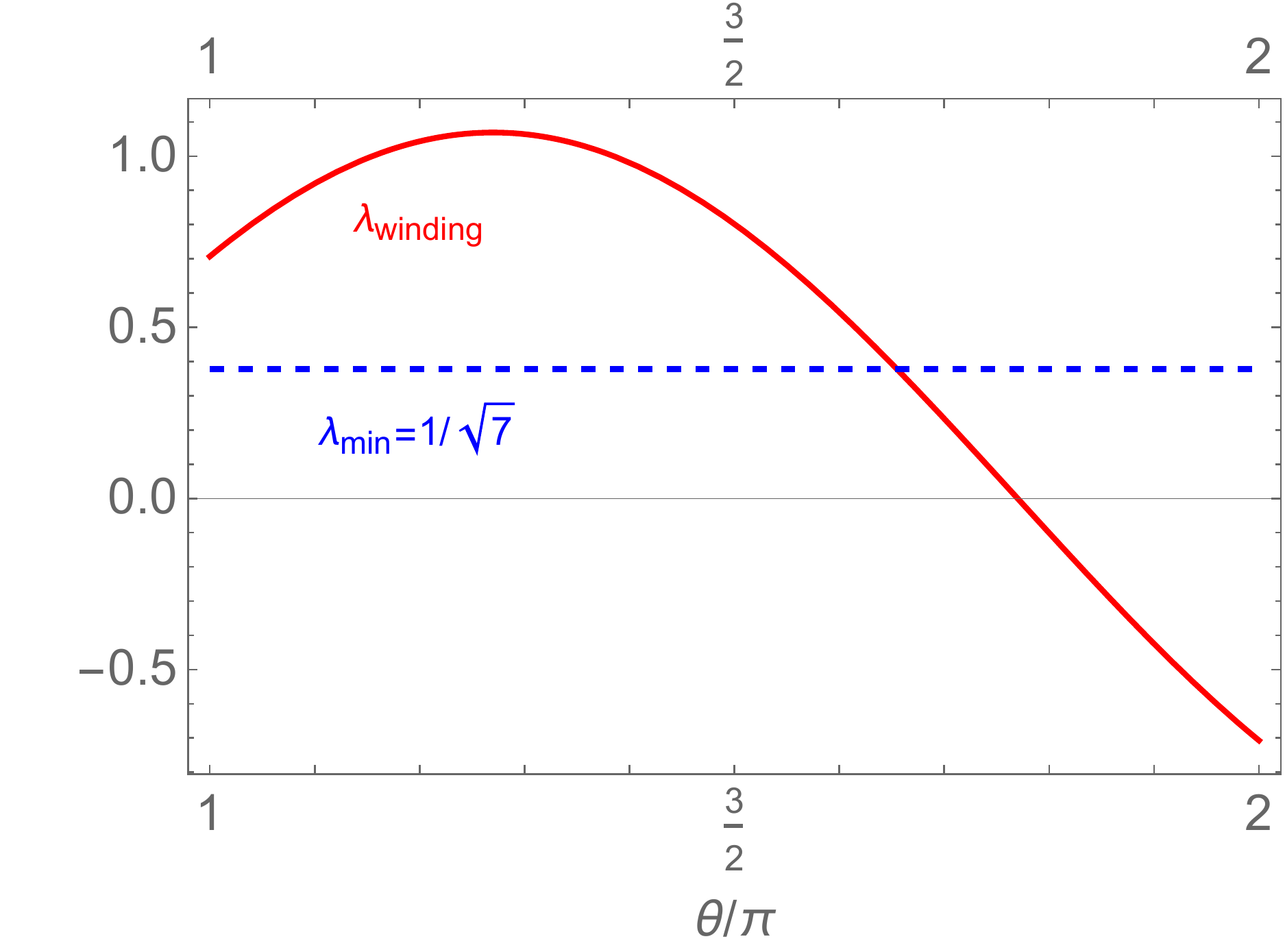}
        \caption{Plots of the numerical coefficient $\lambda$ as a function of $\theta$. Solid red line: The value of $\lambda_\text{winding}$~\eqref{eq:lambda} as a function of $\theta$.
        Dashed blue line: The minimum value of $\lambda$, $\lambda_\text{min}$, dictated by the sharpened distance conjecture.}
        \label{fig:lambda}
    \end{center}
\end{figure}

The value of $\lambda_\text{winding}$~\eqref{eq:lambda} as a function of $\theta$ is plotted in Fig.~\ref{fig:lambda} (solid red line).
The dashed blue line corresponds to the minimum value of $\lambda$, $\lambda_\text{min}$, dictated by the sharpened distance conjecture.
The figure implies that for $\pi<\theta\lesssim1.65\pi$, $\lambda_\text{winding}$ is bigger than $\lambda_\text{min}$.
Since the winding tower whose mass is protected by the BPS condition has a decay rate larger than $\lambda_\text{min}$, the sharpened version of the distance conjecture implies that the above limit must be a 10d decompactification limit, and so the possibility $1$ is correct.

Moreover, from the chiralities of the winding states which we listed earlier, we conclude that any weakly coupled type IIA theory is T-dual to a weakly coupled type IIB theory and vice-versa.
To be more precise, we showed that if quantum gravity is described by IIB/IIA supergravity on a circle with radius $R$ at low energies, then the same theory at sufficiently small (but finite) $R$ is described by IIA/IIB supergravity at sufficiently small energies. 
Note that the 9d theory exists for every value of $R$, even if the 10d description cannot be trusted\footnote{It is expected that the original supergravity description breaks down for $R\to0$ because of the species scale \cite{Dvali:2007wp,Dvali:2007hz} for a string with tension $l_s^{-2}$ is $l_s^{-1}$. Any effective field theory description breaks at $R\ll l_s$.}. 
The statement of duality holds for \textbf{any} radius, and is useful when the radius is sufficiently large in either description.

When the KK/winding state is not the lightest state, the above statement is trivially correct.
However, since we have argued that the KK/winding state is lightest for $R\to0$ or $R\to\infty$, the statement of the T-duality above is non-trivial.

\section{Other dualities in theories with 32 supercharges}\label{sec:duality_32supercharges}
In this section, we argue the relationship between the string dualities of 11d/10d theories (other than IIA/IIB T-duality) with 32 supercharges and the Swampland conjectures.

\subsection{IIA/M-theory} \label{AM}

Now let us study the strong coupling limit of a theory which at low energy is described by the type IIA supergravity. In type IIA, assuming the BPS completeness hypothesis \cite{Kim:2019vuc},\footnote{To be precise, we assume the stronger version of the BPS completeness where the each BPS state is populated by a single particle state. In other words, we assume the existence of the bound state at the threshold~\cite{Sethi:1997pa}.} we have a tower of BPS particles whose masses are protected by supersymmetry.
\begin{align}
m_\text{BPS}\propto \frac{M_{10}}{g^{\frac{3}{4}}},
\end{align}
where $g$ is the coupling constant of the 2-form field in the gravity multiplet (see Eq.~\eqref{cdef}). Upon expressing this mass scale in terms of the canonically normalized spacetime modulus $\hat\phi=\ln(g)/\sqrt{2}$, we find
\begin{align}
    m_\text{BPS}\propto \exp\left(-\frac{3}{\sqrt{8}}\hat\phi\right).
\end{align}
The BPS particles might not be the lightest tower, however, they provide a tower which has a decay rate faster than $\lambda_\text{min}=1/\sqrt{8}$. Therefore, the sharpened version of the distance conjecture implies that this limit cannot be a string limit and must be a decompactification limit. There is a unique field theory in dimension greater than 10 that has 32 superchrages \cite{Nahm:1977tg,Cremmer:1978km,Freedman:2012zz}, which is the 11d supergravity, and the only 1d internal geometry which would preserve all of that supersymmetry is a disjoint union of multiple $S^1$. However, due to the uniqueness of the lower dimensional graviton, the internal geometry must be connected. Therefore, the only possibility is 11d supergravity on $S^1$. Note that since the 11d supergravity does not have any 1-form gauge field, there are no fluxes (in this case Willson line) that can be turned on. 

After dimensional reduction of the 11d supergravity, it is easy to see that the standard matching~\cite{Witten:1995ex} of the fields between M-theory on $S^1$ and type IIA is unique. For example, the 11d supergravity on $S^1$ produces a unique gauge field, which comes from the KK reduction of the metric. Therefore, the particles charged under the type IIA gauge field must have KK momentum. In fact, from the 11d picture, we can also see that the BPS particles must be the KK particles. Therefore, we can match the mass of the 10d BPS particles with the mass of the KK particles to relate the coupling of IIA with the radius in the 11d supergravity picture in a unique way according to the usual string theory argument. 

\subsection{IIB S-duality}

The S-duality of a theory which at low energies is described by type IIB supergravity easily follows from other dualities that we have derived so far. 

In particular, we have shown in section \ref{s2} that such a theory on a circle of radius $R$ is dual to a IIA supergravity on a circle of radius $l_s^2/R$ for sufficiently small $R$. Moreover, in section \ref{AM}, we have shown that type IIA supergravity with $\hat{\phi}\to\infty$ is described by 11d supergravity on a circle. Therefore, type IIB supergravity on a circle has a dual description as an 11d supergravity on a torus. Since the $SL(2,\mathbb{Z})$ symmetry of the Teichmuller parameter of the torus must be a discrete gauge symmetry in the 11d picture, this must also be the case in the type IIB picture for the duality to hold. Therefore, we find that the low-energy $SL(2,\mathbb{Z})$ symmetry of type IIB supergravity cannot be an accidental symmetry, and must indeed be a duality of the theory. 

\section{Dualities in theories with 16 supercharges}
\label{sec:duality_16supercharges}
In this section, we argue the relationship between the string dualities of 10d/9d theories with 16 supercharges and the Swampland conjectures.
\subsection{Duality between \texorpdfstring{Spin$(32)/\mathbb{Z}_2$}{Spin(32)/Z2} and \texorpdfstring{$E_8\times E_8$}{E8*E8} supergravity theories}

Similar to the previous duality, the duality between the Spin$(32)/\mathbb{Z}_2$ and $E_8\times E_8$ supergravity theories follows directly from the sharpened distance conjecture. The key point is that since the gauge group in both theories is maximally enhanced, the charge lattice of the theory on $S^1$ is fully known. In particular, we show that the charge lattice is even and self-dual.

When we compactify the theory on $S^1$, the charge lattice is the direct sum of two lattices, i.e. the weight lattice of the 10d theory and $\Gamma^{1,1}$. The first lattice consists of charges under the KK reduction of 10d gauge fields, and the latter lattice is the charge lattice under $B^{\mu9}$ and $g^{\mu 9}$. A priori, it is non-trivial why the charge lattice must be decomposable this way. However, charges states under $B^{\mu 9}$ are winding strings. The ground state of such winding strings is BPS, and the action of such BPS strings is known from Swampland arguments \cite{Kim:2019ths}. In particular, we know that the gauge symmetry of the spacetime induces a current algebra on the string. Therefore, the winding charge decouples from the other charges. Similarly, the KK charge is known to not mix with the higher dimensional gauge charges. Therefore, the full lattice must decompose into a 2d part and the higher dimensional charge lattice in the absence of Wilson lines. Moreover, given that the charges of the KK charge and the winding charge are known, it is easy to verify that the resulting 2d lattice is even and self-dual, just like the remaining 16d lattice, which is either $e_8^2$ or $d_{16}^+$, where $e_8$ and $d_{16}^+$ are the weight lattices of $E_8$ and Spin$(32)/\mathbb{Z}_2$, respectively. Since the direct sum of two even and self-dual lattices is even and self-dual, the overall charge lattice is even and self-dual. 

All even and self-dual lattices with the signature $(1,17)$ are related by a similarity transformation. Since it is natural to assume that variations of the 9d moduli act by a similarity transformation on the charge lattice \cite{Maharana:1992my,Sen:1994eb} (see Appendix~\ref{massive_vector} for an argument), if an even and self-dual charge lattice is realized only at one point of the 9d moduli, all such lattices can be realized. In particular, if we compactify the $E_8\times E_8$ theory on $S^1$, we can move in the 9d moduli space to change the charge lattice from $\Gamma^{1,1}\oplus e_8^2$ to $\Gamma^{1,1}\oplus d_{16}^+$. Then we can act with a boost of increasing rapidity on $\Gamma^{1,1}$ by moving to an infinite distance of the moduli space. Assuming the BPS completeness hypothesis, there are BPS particles that lie on both axes of $\Gamma^{1,1}$. Since these particles are BPS, we can calculate their masses and see how they depend on the moduli of the 9d theory. Since the answer is unique, we can find the answer via a trick by looking at the dependence of KK towers in the decompactification limits of 10d supergravities on $S^1$. We find that the coefficient of the distance conjecture for such particles is $\sqrt{8/7}$. Given that this number is greater than $\lambda_\text{min}=1/\sqrt{7}$, this limit must decompactify. Moreover, from the charge lattice, we know that the gauge lie group contains Spin$(32)/\mathbb{Z}_2$. Given that the only non-anomalous supergravity with such a gauge algebra is $so(32)$ supergravity, we conclude that $E_8\times E_8$ and Spin$(32)/\mathbb{Z}_2$ supergravities are dual to each other.

\subsection{Duality among heterotic \texorpdfstring{Spin$(32)/\mathbb{Z}_2$}{Spin(32)/Z2}, type I, and type \texorpdfstring{I$^\prime$}{I'}}

Here we consider the strong coupling limit of the Spin$(32)/\mathbb{Z}_2$ supergravity, where the dilaton field in the gravity multiple is taken to infinity. In this limit the supergravity string charged under $B_{\mu\nu}$, which is only BPS state in 10d theories with 16 supercharges, becomes infinitely heavy. Therefore, the tower of the light states must be non-BPS.
This is exactly what we know from string theory~\cite{Polchinski:1995df}.
This limit corresponds to the perturbative type I theory where a non-supersymmetric type I fundamental string becomes tensionless. The question is how to arrive at the same conclusion in a bottom-up manner.

From the sharpened distance conjecture, the strong coupling limit is either the tensionless string or the decompactification limit.
In the latter case, the theory must decompactify to the 11d supergravity.
In the following, we argue that there are no compactifications of 11d supergravity which leads to the 10d Spin$(32)/\mathbb{Z}_2$ supergravity.
This indicates that the strong coupling limit is the tensionless non-BPS string limit.

\begin{figure}
    \centering

\tikzset{every picture/.style={line width=0.75pt}} 

\begin{tikzpicture}[x=0.75pt,y=0.75pt,yscale=-1,xscale=1]

\draw    (228.85,58.61) -- (370.89,59.43) ;
\draw    (370.89,59.43) .. controls (410.63,-3.67) and (442.95,34.91) .. (443.53,58.92) .. controls (444.1,82.93) and (411.26,130.36) .. (370.89,59.43) -- cycle ;
\draw  [fill={rgb, 255:red, 208; green, 2; blue, 27 }  ,fill opacity=1 ] (368.56,57.05) .. controls (369.53,55.54) and (371.54,55.1) .. (373.05,56.07) .. controls (374.56,57.04) and (375,59.05) .. (374.03,60.56) .. controls (373.06,62.07) and (371.05,62.51) .. (369.54,61.54) .. controls (368.03,60.57) and (367.59,58.56) .. (368.56,57.05) -- cycle ;
\draw  [fill={rgb, 255:red, 208; green, 2; blue, 27 }  ,fill opacity=1 ] (226.11,56.86) .. controls (227.08,55.34) and (229.09,54.91) .. (230.6,55.87) .. controls (232.11,56.84) and (232.55,58.85) .. (231.58,60.37) .. controls (230.61,61.88) and (228.6,62.32) .. (227.09,61.35) .. controls (225.58,60.38) and (225.14,58.37) .. (226.11,56.86) -- cycle ;
\draw  [fill={rgb, 255:red, 208; green, 2; blue, 27 }  ,fill opacity=1 ] (262.89,57.87) .. controls (263.85,56.36) and (265.86,55.92) .. (267.38,56.89) .. controls (268.89,57.86) and (269.33,59.87) .. (268.36,61.38) .. controls (267.39,62.89) and (265.38,63.33) .. (263.87,62.36) .. controls (262.35,61.39) and (261.92,59.38) .. (262.89,57.87) -- cycle ;

\end{tikzpicture}
    \caption{A hypothetical exotic internal geometry for 11d supergravity. The consistency of such compactification depends on the existence of appropriate non-perturbative branes sitting at the vertices.}
    \label{ExoticM}
\end{figure}
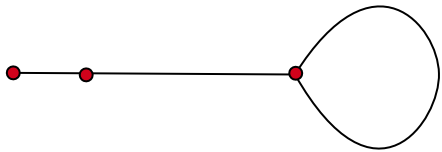

To this end, suppose the strong coupling limit of the 10d supergravity decompactifies to the 11d supegravity.
Since the 11d supergravity has 32 supercharges, the internal dimension must have features that break the half of the supersymmetry.
A priori, one can imagine any 1d graph with singular points as a viable option of the compactification (see Figure \ref{ExoticM})\footnote{The singular points are viewed as positions of the brane.}. 

However, from the Swampland conjectures, it was argued in Ref.~\cite{Bedroya2023} that the only background preserving half of the supersymmetry is an interval. To see why we first put the 10d theory on a circle to find a type IIA background on the corresponding 1d graph. 
Then, we introduce a BPS 4-brane coupled with a 3-form field. 
The BPS 4-brane is a supergravity solution, and is required from the BPS completeness hypothesis.
In Ref.~\cite{Bedroya2023}, it was pointed out that since the BPS 4-brane solutions placed at a point in the graph must be continuously transformable to zero-sized gauge theory instantons~\cite{Witten:1995gx} of the 9d theory, no-global symmetry conjecture implies that they correspond to different points in the Coulomb branch of a BPS 4-brane in 9d which we can call a small instanton.\footnote{The small instanton was used to classify the possible patterns of the gauge algebra of supergravity theories~\cite{Hamada:2021bbz,Bedroya:2021fbu}.}

The worldvolume theory of small instanton at a generic point of its 1d Coulomb branch is simply a BPS 4-brane placed on the graph.\footnote{Moving the internal direction is identified as Coulomb branch as it is real one dimension. We need complex scalars to obtain the Higgs branch.} 
Therefore, the internal geometry must match the Coulomb branch. However, the theory on large scales is described by a 5d SCFT and the vertices of the internal geometry are the Coulomb branch singularities of such 5d SCFT. Given the classification of such theories and their corresponding singularities, we know that the only allowed vertices are the end of the interval points. Therefore, a 1d graph such as Figure \ref{ExoticM} does not preserve half of the supersymmetry.
The only allowed geometry is an interval. 

Now we have found that the only way to obtain a 10d theory with $16$ supercharges is to compactify the 11d supergravity on the interval.
In this case, it is well-known that the anomaly cancellation fixes the gauge group \cite{Horava:1995qa} to be $E_8\times E_8$. 
Therefore, we conclude that the strong coupling limit of the Spin$(32)/\mathbb{Z}_2$ supergravity cannot be a KK limit, and according to the sharpened distance conjecture, it must be a string limit. As we pointed out earlier, it follows that the corresponding fundamental string is non-supersymmetric, as expected from the type I string theory. 

So far, we have explained the duality between the Spin$(32)/\mathbb{Z}_2$ theory and type I by showing that Swampland conditions imply the existence of the type I string. The duality between the type I and type I$^\prime$ theory, on the other hand, is more subtle and was the subject of a separate recent paper \cite{Bedroya2023}. We review that argument in Appendix \ref{IIp}. Using various Swampland conjectures, authors showed that the strong coupling limit of the 9d supergravity is always described by type IIA supergravity on an interval. Moreover, there are BPS domain walls along the interval whose location corresponds to the 9d moduli and uniquely determine the unbroken gauge symmetry via Swampland argument.

\subsection{\texorpdfstring{$E_8\times E_8$}{E8*E8} and 11d supergravity}

Now we combine the chains of dualities that we have established to identify the strong coupling limit of the $E_8\times E_8$ theory.

In Appendix \ref{IIp}, we review the Swampland argument of Ref.~\cite{Bedroya2023} which shows that any 9d supergravity has a strong coupling limit which decompactifies to type I$^\prime$ background (type IIA on an interval). 
We can apply this knowledge to the $E_8\times E_8$ supergravity on $S^1$ and take its strong coupling limit of $\phi\rightarrow\infty$.
According to the type IIA picture, this is the limit where the 10d coupling of type IIA is taken to infinity. From the discussion in Section~\ref{AM}, the theory decompactifies to 11d supergravity on $S^1\times I^1$.
Therefore, we find that the strong coupling limit of the $E_8\times E_8$ is 11d supergravity on $I^1$.

\subsection{Chain of dualities and 9d Supergravity}

Our bottom-up arguments for the dualities between the 10d and 11d supergravities have direct implications for lower dimensions. We will argue below, that in 9d supergravities with 17 vector multiples, we know the structure of almost all infinite distance limits.

In Ref.~\cite{Bedroya2023} it was shown that any strong coupling limit of a 9d $\mathcal{N}=1$ supergravity is type IIA on an interval (type I$^\prime$ theory). But now that we have a complete web of dualities between the higher dimensional theories, we know that the type I$^\prime$ theory shares its moduli space with other 9d supergravities such as type I on a circle, Heterotic on a circle, or M-theory on a cylinder. 
Therefore, we can tell that there are corners of the 9d supergravity with the rank 17 that decompactify to each one of those theories. 
But we can do even better! Our proof of dualities in the previous sections not only demonstrate that the mentioned theories share a mutual moduli space, but also explains how the moduli of the theories are connected. Therefore, we know exactly how the moduli between any two such descriptions are related. For example, the duality between type IIA and 11d supergravity on circle, tells us how the moduli of type I$^\prime$ are related to the moduli of 11d supergravity on a cylinder. Therefore, if an infinite distance limit is the type I$^\prime$ supergravity, we can use that matching of the moduli to figure out what limits go to 11d supergravity on the cylinder. In 9d theories of maximal rank ($r=17$), all the infinite distance limits are already covered by the dualities we derived. Therefore, our arguments can be used as a bottom-up derivation that the infinite distance limits of the 9d theory must behave exactly as superstring theory predicts \cite{Aharony:2007du}.

Note that the reason we can identify the infinite distance limits is that a measure one subset of infinite distance limits decompactify to higher dimensions. This result was shown in \cite{Bedroya2023} using the sharpened distance conjecture and the finiteness of the black hole entropy. This is a very strong result which demonstrates how non-trivial sharpened distance conjecture is. As long as we identify only one decompactification limit precisely, we can use the proven dualities to infer the other infinite distance limits. 

\section{Conclusions}\label{sec:conclusions}

\begin{figure}
    \centering

\tikzset{every picture/.style={line width=0.75pt}} 

\begin{tikzpicture}[x=0.75pt,y=0.75pt,yscale=-1,xscale=1]

        \node[draw,outer sep=0pt] (a) at (100,0) {IIA};
        \node[draw,outer sep=0pt]  (b) at (0,0) {IIB};
        \node[draw,outer sep=0pt]  (c) at (150,-100) {11d SUGRA};
        \node[draw,outer sep=0pt,font=\fontsize{10pt}{.3cm}\selectfont]  (d) at (50,100) {9d SUGRA 32Q};
        \node[draw,outer sep=0pt]  (e) at (200,0) {$E_8\times E_8$};
        \node[draw,outer sep=0pt]  (f) at (300,0) {$SO(32)$};
        \node[draw,outer sep=0pt,font=\fontsize{10pt}{.3cm}\selectfont]  (g) at (250,100) {9d SUGRA 16Q};
        \draw[arrows=|->] (c) -- (a);
        \draw[arrows=|->] (a) -- (d); 
        \draw[arrows=|->] (a) -- (g); 
        \draw[arrows=|->] (b) -- (d); 
        \draw[arrows=|->] (c) -- (e); 
        \draw[arrows=|->] (e) -- (g);
        \draw[arrows=|->] (f) -- (g);

\draw (45,-70) node [anchor=north west,align=right,font=\fontsize{9pt}{.3cm}\selectfont
][inner sep=0.5pt]    {\textcolor{red}{Stronger BPS}\\ \textcolor{red}{completeness}};
\draw (20,120) node [anchor=north west,font=\fontsize{10pt}{.3cm}\selectfont
][inner sep=0.5pt]    {T-duality};
\draw (0,140) node [anchor=north west,font=\fontsize{9pt}{.3cm}\selectfont
][inner sep=0.5pt]    {\textcolor{red}{BPS completeness}};
\draw (175,-70) node [anchor=north west,align=right,font=\fontsize{9pt}{.3cm}\selectfont
][inner sep=0.5pt]    {\textcolor{red}{Ref.\cite{Bedroya2023}+Stronger}\\ \textcolor{red}{BPS completeness}};
\draw (280,-50) node [anchor=north west,font=\fontsize{10pt}{.3cm}\selectfont
][inner sep=0.5pt]    {S-duality};
\draw (260,-37) node [anchor=north west,align=right,font=\fontsize{9pt}{.3cm}\selectfont
][inner sep=0.5pt]    {\textcolor{red}{4-brane Coulomb}\\
\textcolor{red}{=internal space}};
\draw (220,120) node [anchor=north west,font=\fontsize{10pt}{.3cm}\selectfont
][inner sep=0.5pt]    {T-duality};
\draw (170,140) node [anchor=north west,align=right,font=\fontsize{9pt}{.3cm}\selectfont
][inner sep=0.5pt]    {\textcolor{red}{9d moduli space=charge lattice}};
\draw (100,-40) node [anchor=north west,font=\fontsize{10pt}{.3cm}\selectfont
][inner sep=0.5pt]    {$S^1$};
\draw (45,50) node [anchor=north west,font=\fontsize{10pt}{.3cm}\selectfont
][inner sep=0.5pt]    {$S^1$};
\draw (245,50) node [anchor=north west,font=\fontsize{10pt}{.3cm}\selectfont
][inner sep=0.5pt]    {$S^1$};
\draw (190,-40) node [anchor=north west,font=\fontsize{10pt}{.3cm}\selectfont
][inner sep=0.5pt]    {$S^1/\mathbb{Z}_2$};
\draw (150,55) node [anchor=north west,font=\fontsize{10pt}{.3cm}\selectfont
][inner sep=0.5pt]    {$S^1/\mathbb{Z}_2$};
\draw (150,75) node [anchor=north west,align=right,font=\fontsize{9pt}{.3cm}\selectfont
][inner sep=0.5pt]    {\textcolor{red}{Ref.\cite{Bedroya2023}}};

\end{tikzpicture}
    \caption{Summary of our result. The upper, middle, and lower boxes correspond to 11d, 10d, and 9d supersymmetric theories, respectively. The relation among them and dualities are denoted by the arrows or texts in black color. The assumptions we used to derive the dualities are written in red color. The sharpened distance conjecture are generically used, and is not written in the figure.}
    \label{fig:Summary}
\end{figure}
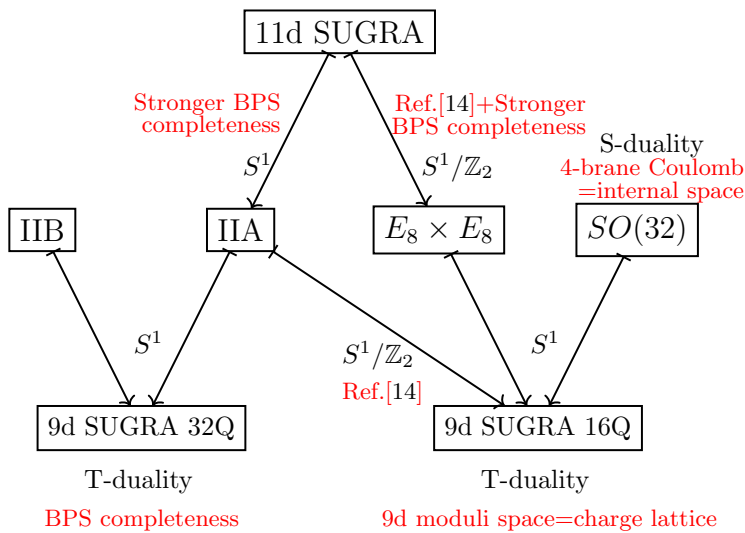

Swampland principles were previously used to argue that the only consistent  10d and 11d supergravities are those appearing in string theory \cite{Kim:2019vuc}. However, the dualities between the different theories were less explored. In fact, the universality of dualities was the main motivation behind the Swampland distance conjecture which was later sharpened in subsequent formulations. We provided a concrete argument for this relationship by showing that not only are Swampland conjectures motivated by string theory dualities, but also string dualities can be derived from Swampland conjectures under several assumptions.
Our results are summarized in Figure~\ref{fig:Summary}.

This work demonstrates the power of the Swampland conjectures and allows us to recast the main lessons from string theory and its universality in terms of Swampland conjectures. 

\section*{Acknowledgement}
We thank Miguel Montero, Houri Christina Tarazi, and Cumrun Vafa  for many insightful discussions. YH would like to thank Harvard University and University of Wisconsin-Madison for their hospitality during part of this work. The work of AB is supported by a grant from the Simons Foundation (602883, CV) and by the NSF grant PHY-2013858. The work of YH is supported in part by MEXT Leading Initiative for Excellent Young Researchers Grant Number JPMXS0320210099.

\appendix

\section{Tension of the supergravity string}\label{A1}

We use supersymmetry to find the tension of a BPS supergravity string. We present the calculation in 7d where there is a nice way of labeling the vectors in the gravity multiplet. However, as we will explain later, the argument applies to any theory with 16 supercharges.

The gravity multiplet in $7d$ has three graviphotons and the smallest irreducible spinor representation is a pseudo Majorana spinor which can be viewed as two sets of 8 real grassman numbers. These two are actually related. The gravity multiplet can be packaged in representations of $Sp(1)$. The pseudo Majorana spinors carry a 2 dimensional representation and are labeled by $i\in\{1,2\}$ and the graviphotons furnish a three dimensional representation $A^{ij}$ where $A$ is antisymmetric in $i$ and $j$. The theory also has a pseudo Majorana spinor $\chi^i$ which can be viewed as a doublet in $Sp(1)$. 

In this notation, the supersymmetry transformation rules take a simple form.  The fields are tetrads $e^\mu_A$, gravitino $\psi^\mu_{a,i}$, graviphotons $A^{\mu i}_j$, 2-form $B^{\mu\nu}$, fermions $\chi_i^a$, and the dilaton $\phi$. Greek indices are spacetime indices, $\{A,B,...\}$ are Lorentz indices, $\{i,j,...\}$ are $Sp(1)$ indices, and $\{a,b,...\}$ are spinor indices. We will often drop the spinor indices in the calculations. Also, $\gamma^\mu=\gamma^Ae_A^\mu$ where $\gamma^A$ are 7d Dirac matrices. 

Now let us go back to our question of interest. Consider the string electrically coupled to $B^{\mu\nu}$. This string has an action term
\begin{align}\label{Baction}
    S\propto i\int \star B\wedge dX\wedge dX+...
\end{align}

Now we consider the action of supersymmetry on the field $B_{\mu\nu}$ to see how this action transforms. The full list of possible supersymmetry transformations of the fields in the gravity multiplet is given as below \cite{Han:1985ku}\footnote{We have used the fact that the spinors are pseudo Majorana to rewrite the terms in a slightly different way that makes the calculations simpler. In particular, we have expressed the transformation rules of all of the bosonic fields in terms of $\bar\epsilon$ rather than $\epsilon$.}.

\begin{align}\label{SUSYrules}
    \delta e^A&=\kappa\bar\epsilon^i\gamma^A\wedge \psi_i\nonumber\\
    \delta\psi_i&=\frac{2}{\kappa}D\epsilon_i+c_1 \star[\star(\gamma\wedge\gamma\wedge\gamma \epsilon_j)\wedge F^i_j]e^{q\kappa\phi}\nonumber\\
    &+d_1\star(\gamma\epsilon_j\wedge\star F^j_i)e^{q\kappa\phi}\nonumber\\
    &+c_2\star[\star(\gamma\wedge\gamma\wedge\gamma\wedge\gamma\epsilon_i)\wedge G]e^{q\kappa\phi}\nonumber\\
    &+d_2\star(\gamma\wedge\gamma\epsilon_i\wedge\star G)e^{r\kappa\phi}+\text{bilinear fermions}\nonumber\\
    \delta\chi_i&=c_3\star[(\gamma\wedge\gamma\epsilon_i)\wedge\star F^i_j]e^{q\kappa\phi}\nonumber\\
    &+c_4\star[(\gamma\wedge\gamma\wedge\gamma\epsilon_i)\wedge\star G]e^{r\kappa\phi}\nonumber\\
    &+c_5\star(\gamma\epsilon_i\wedge\star d\phi)+\text{bilinear ferimoins}\nonumber\\
    \delta A^j_i&=f_1(\bar\epsilon^j\psi-\frac{1}{2}\delta^{j}_{i}\bar\epsilon^k\psi)e^{-q\kappa\phi}\nonumber\\
    &+f_3(\bar\epsilon^j\gamma\chi_i-\frac{1}{2}\delta^{j}_{i}\bar\epsilon^k\gamma\chi_k)e^{-q\kappa\phi}\nonumber\\
    \delta B&=(f_2\bar\epsilon^i\gamma\wedge\psi_i+f_4\bar\epsilon^i\gamma\wedge\gamma\chi_i)e^{-r\kappa\phi}+p_2A^j_i\wedge\delta A^i_j\nonumber\\
    \delta\phi&=f_5\bar\epsilon^i\chi_i,
\end{align}
where $G=dB$, $F^{i}_j=dA^i_j$, $\gamma^A$ is a 0-form, $\gamma$ is a 1-form, $\psi_i$ is a 1-form, $\kappa$ is proportional to the 7d Newton constant, and $\{c_i,d_i,f_i,p_i,r,q\}$ are all non-zero numerical coefficients that are determined by the closure of the supersymmetry algebra. Note that, modulo the $Sp(1)$ index $\{i,j\}$, the above expressions hold for all dimensions. Note that the exponential factor $e^{-r\kappa\phi}$ is the fundamental charge of the supergravity string, because the kinetic term for $B$ takes the following form
\begin{align}
    S^{7d}_B=-\int\frac{1}{2}dB\wedge\star dBe^{2r\kappa\phi}
\end{align}
The supersymmetry transformation of the action \eqref{Baction} is $ \delta B=(f_2\bar\epsilon^i\gamma\wedge\psi_i+f_4\bar\epsilon^i\gamma\wedge\gamma\chi_i)e^{-r\kappa\phi_0}+p_2A^j_i\wedge\delta A^i_j+\hdots$. For the string action to be supersymmetric, this variation must be canceled by the supersymmetry transformation of other terms in the action. Note that $\phi_0$ is the asymptotic value of $\phi$ which sets the coupling constant of the theory. It is almost clear how to cancel the last term since it already contains the supersymmetry transformation of $A^j_i\wedge A^i_j$. What is less clear, is how to cancel the other two terms. We will focus on $\bar\epsilon^i\gamma\wedge\psi_ie^{-r\kappa\phi_0}$ term. It is easy to see that if the supersymmetric variation of a term does not depend on anything other than $e$ and $\psi$ and has a single $\gamma$ matrix, that term must only depend on $e$.

Therefore, we are looking for a geometric 2-form. Moreover, our 2-form must be a function of $e$ and not its derivatives. The only such area form is the induced volume form on the worldsheet. Let us verify that the induced area gives the correct variation under supersymmetry transformation. Suppose $h$ is the induced metric. We can express $h_{\alpha\beta}$ in terms of the tetrads as
\begin{align}
    h_{\alpha\beta}=\partial_\alpha X^{\mu}\partial_\beta X^{\nu}e_\mu^M e_\nu^N \eta_{MN}.
\end{align}
Moreover, we can write the determinant $h$ as
\begin{align}
    h=\frac{1}{2}h_{\alpha\beta}h_{\alpha'\beta'}\epsilon^{\alpha\alpha'}\epsilon^{\beta\beta'}.
\end{align}
and its supersymmetry variation is 
\begin{align}
    \delta \sqrt{-h}=-\frac{h_{\alpha'\beta'}\epsilon^{\alpha\alpha'}\epsilon^{\beta\beta'}}{\sqrt{-h}}\partial_\alpha X^{\mu}\partial_\beta X^{nu}e_{\nu}^N\delta e_\mu^M\eta_{MN}.
\end{align}
In conformal coordinates, the first term simplifies as
\begin{align}
    \frac{h_{\alpha'\beta'}\epsilon^{\alpha\alpha'}\epsilon^{\beta\beta'}}{\sqrt{-h}}=\epsilon^{\alpha\beta}.
\end{align}
Working in conformal coordinates, and plugging in $\delta e_\mu^M$ from \eqref{SUSYrules} gives

\begin{align}
   \delta\sqrt{-h}d^2\sigma=-e_\nu^j\delta e_\mu^i\eta_{ij}dx^\mu dx^\nu=-\kappa \bar\epsilon\wedge\gamma\psi.
\end{align}
which is coordinate independent and holds in every coordinate system. Therefore, we find that to cancel the supersymmetry transformation of \eqref{Baction} we must add 
\begin{align}
    i\frac{f_2}{\kappa}\int\sqrt{-h}e^{-r\kappa\phi}.
\end{align}
The prefactor $\frac{f_2}{\kappa}$ is a pure imaginary number that is determined by the closure of supersymmetry algebra. Similarly, the number $r$ is determined by supersymmetry. For example, in 7d, we have 
\begin{align}
    f_2=\frac{i}{\sqrt{2}},~~r=-\frac{2}{\sqrt{5}}.
\end{align}
In general, we find that supersymmetry dictates the term
\begin{align}
   S_\text{Nambu-Goto}= -\frac{A_d}{2\pi}\int \sqrt{-h} e^{c_d\phi}
\end{align}
for some dimension dependent positive constants $A_d$
 and $c_d$. This also implies that the tension of the supergravity string is given by
 \begin{align}
    T=A_d e^{c_d\phi_0},
 \end{align}
 where $e^{c_d\phi_0}$ is the coupling of the 2-form $B$ in the gravity multiplet.

In 10 dimensions, $r=1/\sqrt{2}$. Therefore, the tension of the string is given by
 \begin{align}
    T=A e^{\hat\phi_0/\sqrt{2}},
 \end{align}
for some constant $A$. We can simplify our equations by shifting $\hat\phi$ by a constant to absorb the coefficient $A$. This shift will make the normalization of $B$ non-canonical, but the normalization of $\hat\phi$ will remain canonical. 

\begin{align}
    T=e^{\hat\phi_0/\sqrt{2}}M_{10}^2.
\end{align}

Note that in our convention, 
\begin{align}
    S=\frac{T}{2\pi}\cdot\text{Area}.
\end{align}

\section{Ground states of winding BPS string}\label{A2}

In this section, we determine the Spin$(8)$ representation of the ground state of the static winding BPS supergravity string\footnote{The partial supersymmetry breaking by the winding string was discussed, e.g., in Refs.~\cite{Hughes:1986dn,Dabholkar:1990yf}.}.
After fixing the coordinates on the worldsheet, the low energy action on the string would be in terms of spacetime coordinates $X^i$ with $i\in\{1,2,\hdots,8\}$. In Appendix A, we found that the action around the ground state of a BPS supergravity string is 
\begin{align}\label{BA}
    S= -T\int \sqrt{- g} d^2\,\sigma.
\end{align}
where $T=A_de^{c_d\phi}$ and $g$ is the induced metric on the string worldsheet. The above action is trustable for perturbations around BPS configurations as well. 

The action \eqref{BA} is not the full worldsheet action. Any spacetime supercharge that is preserved by the string, maps to a global charge on the worldsheet. The corresponding worldsheet charges can be found by the Green-Schwarz formalism~\cite{Green:1981yb} for studying supersymmetric branes \cite{Green:1983wt}.
However, the detail of the worldsheet action is not important for us.

We compactify the $X^9$ direction with the radius $R$, $X^9\in [0,2\pi R)$.
We consider a static string winding around a circle. 
We choose the worldsheet coordinates as 
\begin{align}
    (\sigma,\tau)=\left(\frac{X^9}{R},X^0\right),
\end{align}
and we have $g_{\ta\tau}=1$ and $g_{\sigma\sigma}=R^2$. 
This gauge choice is known as the static gauge or unitary gauge~\cite{Luscher:2004ib,Aharony:2009gg,Aharony:2013ipa}.

The winding string along the $X^9$ direction breaks the Lorentz symmetry as
\begin{align}
    SO(9,1)\to SO(8)\times SO(1,1).
\end{align}
 
Let us take a closer look at the supercharges in 10d non-compact spacetime and their action on the BPS string. Suppose we start with a supergravity with 32 supercharges. The BPS string preserves half of the supersymmetry. These supercharges act on the worldsheet fields and must be in fermionic representations of $SO(8)$ corresponding to rotations in the transverse coordinates to the string. This symmetry manifests itself as the R-symmetry of the worldsheet theory. The smallest spinor representation of $SO(8)$ is 8 dimensional. The supercharges must also furnish representations of $SO(1,1)$ which is the Lorentz group of the worldsheet. The irreducible representations of $SO(1,1)$ are one-dimensional and can be left or right handed. However, since the R-symmetry maps the supercharges in an irreducible representation of $SO(8)$ to each other, they must all have the same worldsheet handedness. Therefore, the supercharges lead to worldsheet charges that come in groups of $8$ that all have the same worldsheet handedness and are in the vector representation of $SO(8)$. For theories with 32 spacetime supercharges, we find the following two possibilities:

We consider the dimensionally reduced 9d supersymmetry algebra.
In Spin$(8)$ notation corresponding to the spatial rotation, we have~\cite{Green:1987sp}\footnote{The minimal spinor in 9d is 16 components Majorana spinor, which can be decomposed into $\mathbf{8}_s+\mathbf{8}_c$ of Spin$(8)$.
Since we are working in the theories with $32$ supercharges, we have two copies of the Majorana spinor.}
\begin{align}
    \{Q^a_A,Q^{\dot{b}}_B\}\sim\gamma_{a\dot{a}}^ip^i\delta_{AB},
\end{align}
where $i=1,\cdots,8$ is the direction transverse to the string, $A,B=1,2$ is the label of the supersymmetry charges, and $a=1,\cdots,8$ and $\dot{a}=1,\cdots,8$ are the indexes for $\mathbf{8}_S$ and $\mathbf{8}_C$ representations, respectively.
The string states are not invariant under $p^i$ action since this generates  translation along the transverse direction to the string.
This indicates that the BPS string can preserve either $Q^a$ or $Q^{\dot{a}}$, but not both.

Suppose that $Q^a$ is preserved and $Q^{\dot{a}}$ is broken. Then, we can normalize $Q^{\dot{a}}$ in such a way that
\begin{align}
    \{Q^{\dot{a}},Q^{\dot{b}}\}= \delta^{\dot{a}\dot{b}},
\end{align}
is satisfied. This is analog to the Clifford algebra, and by utilizing the 
triality relation in Spin$(8)$, we see that the representation of the algebra above is
\begin{align}
    \mathbf{8}_v\oplus\mathbf{8}_s.
\end{align}
Similarly, if $Q^{\dot{a}}$ is preserved and $Q^a$ is broken, then the algebra of the broken supersymmetry tells us that the representation is
\begin{align}
    \mathbf{8}_v\oplus\mathbf{8}_c.
\end{align}
Therefore, by looking at the broken supersymmetry, we can find the Spin$(8)$ representation of the BPS supergravity string.
\begin{itemize}
    \item 10d $\mathcal{N}=(1,1)$ supergravity (IIA supergravity): 

    The supercharges are the two 10d Majorana-Weyl spinors with different chirality.
    These supercharges are decomposed as
    \begin{align}
    &
    (\mathbf{8}_s,1/2)\oplus(\mathbf{8}_c,-1/2),
    &&\text{and}
    &&
    (\mathbf{8}_s,-1/2)\oplus(\mathbf{8}_c,1/2),
    \end{align}
    under $SO(8)\times SO(1,1)$. Here $-1/2$ and $1/2$ correspond to the left and right-handed fermions, respectively. 

    Depending on the symmetry breaking pattern, we obtain either $\mathcal{N}=(0,16)$ or the vector $\mathcal{N}=(8,8)$ as a 2d worldsheet theory.
    In the following, we argue that it is not possible to obtain the 2d $\mathcal{N}=(0,16)$ worldsheet theory assuming that the theory flows to an SCFT without the symmetry enhancement.

    First, as the perturbative anomaly of the bulk theory is canceled without Green-Schwarz mechanism~\cite{Green:1984sg}, there is no anomaly inflow~\cite{Callan:1984sa} to the BPS string.
    This indicates that the central charges of the left-mover and right-mover satisfy the relation~\cite{Kim:2019ths}
    \begin{align}
        c_L-c_R=0.
    \label{eq:gravitational_anomaly}\end{align}
    Moreover, we use the knowledge of $\mathcal{N}=(0,2)$ superconformal subalgebra in the $\mathcal{N}=(0,16)$ theory. By identifying the R-symmetry $U(1)_R$ of the $\mathcal{N}=(0,2)$ superconformal algebra as an $SO(2)$ subgroup of the $SO(8)$ rotation group, we obtain the relation $c_R=0$~\cite{Kim:2019ths}.
    In total, we have
    \begin{align}
        c_L=c_R=0.
    \label{eq:R_anomaly}\end{align}
    However, this contradicts the fact that there must be the center of mass degrees of freedom.

    Therefore, we conclude that the worldsheet theory possesses $\mathcal{N}=(8,8)$.\footnote{In this case, Eq.~\eqref{eq:gravitational_anomaly} is still correct, but Eq.~\eqref{eq:R_anomaly} is modified.} The supercharge preserved by the BPS string is either
    \begin{align}
    &
    (\mathbf{8}_s,1/2),
    &&\text{and}
    &&
    (\mathbf{8}_s,-1/2),
    \end{align}
    or
    \begin{align}
    &
    (\mathbf{8}_c,-1/2),
    &&\text{and}
    &&
    (\mathbf{8}_c,1/2).
    \end{align}    
    The worldsheet supersymmetry is $\mathcal{N}=(8_s,8_s)$ or $(8_c,8_c)$.
    In both cases, the Spin$(8)$ chirality is the same.
    This means that the Spin$(8)$ representation of the BPS string is
    \begin{align}
        (\mathbf{8}_v\oplus\mathbf{8}_{s(c)})\otimes(\mathbf{8}_v\oplus\mathbf{8}_{s(c)}).
    \end{align}
    Note that BPS states contain $2^8$ states while non-BPS states contain $2^{16}$ states.
    
    \item 10d $\mathcal{N}=(2,0)$ supergravity (IIB supergravity): 
    
    The supercharges are the two 10d Majorana-Weyl spinors with the same  chirality.
    Both supercharges are decomposed as
    \begin{align}
    Q_{A=1,2}:(\mathbf{8}_s,1/2)\oplus(\mathbf{8}_c,-1/2),
    \end{align}
    under $SO(8)\times SO(1,1)$.

    As in the previous case, the supercharge preserved by the BPS string is
    \begin{align}
        &Q_{A=1}:(\mathbf{8}_s,1/2),
        &&Q_{A=2}:(\mathbf{8}_c,-1/2),
    \end{align}
    or vice versa.

    Therefore the worldsheet supersymmetry is $\mathcal{N}=(8_s,8_c)$ or $(8_c,8_s)$, and the Spin$(8)$ representation of the BPS string is
    \begin{align}
        (\mathbf{8}_v\oplus\mathbf{8}_{s(c)})\otimes(\mathbf{8}_v\oplus\mathbf{8}_{c(s)}).
    \end{align}
    
\end{itemize}

\section{Type I and type \texorpdfstring{I$^\prime$}{I'}}\label{IIp}

The duality between the type I and type I$^\prime$ theories is tricky and has been recently argued using Swampland principles in Ref.~\cite{Bedroya2023}. We review the argument here. 

Let us compactify the $SO(32)$ supergravity on a circle and take the small radius $R$ limit while taking the dilaton $\phi$ to infinity. We can choose different relative rates for them,
\begin{align}
\hat\phi=\alpha\ln(l_{10}/R).
\end{align}
where $\hat\phi$ is the canonically normalized dilaton. Since the dilaton goes to infinity, we can think of this theory as type I supergravity on a circle of radius $R$. 

Since $\alpha$ takes value in an uncountable set and the number of string limits is countable (see Ref.~\cite{Bedroya2023}), almost every such limit must decompactify. Therefore, we have a strong coupling limit of the 9d supergravity, which must decompactify. Given that all the BPS particles remain massive, the number of options is very limited. For example, this limit cannot decompactify to an $\mathcal{N}=1$ supergravity on the circle, because in that case the KK tower would be a light BPS tower. In Ref.~\cite{Bedroya2023}, this observation was used along with the classification of 5d SCFTs to show that for large enough $\alpha$, the only possible decompactification is the type IIA supergravity on an interval with some BPS domain walls along the interval. As reviewed in \cite{Bedroya2023}, the moduli of the 9d supergravity can be exactly matched to the location of the domain walls along the interval, and the gauge theory living on the domain walls can be identified without relying on string theory. This is the type I$^\prime$ theory.

\section{Massive vector multiplets in supergravity}\label{massive_vector}

In this appendix, we argue that $N=16$ supersymmetry determines the transformation of the charge lattice under variation of the moduli. The idea is to show that the coupling of any massive particle to massless vectors depends on the scalars in the same vector multiplet in a specific way due to supersymmetry. Since these couplings set the charges, the charges are set by the spacetime moduli. 

First, let us remind ourselves of the supermultiplets in theories with 16 supercharges and dimensions greater than $6$. The only massless multiplets are the gravity multiplet and the vector multiplet. However, we can also have massive multiplets. For example, if we start with a theory with a non-Abelian gauge symmetry and move in the Coulomb branch, we can Higgs some of the previously massless vector multiplets into massive vector multiplets. From the Higgsing argument, it is easy to see that the massive vector multiplets always have $9-d$ scalars in $d$ dimensions. For example, in 9 dimensions, the massless vector multiplet has one real scalar, which gets absorbed into the vector, turning it into a massive vector. The same thing happens in all other dimensions in $N=16$ theories. 

If the mass of the massive vector field is sufficiently small, we must be able to incorporate it into the field theory. In order to get some intuition about the supersymmetric coupling of the massive vector multiplet to massless multiplets, let us start with the example of a Higgsed gauge group.

It is helpful to work with the $O(10-d,k)$ formulation of supergravities \cite{Hohm:2014sxa} where $d$ is the dimension of spacetime and $k$ is the dimension of the gauge group. We use the convention where the $(10-d,k)$ metric is 
\begin{align}
\eta=\begin{pmatrix}0&\mathds{1}_{10-d}&0\\\mathds{1}_{10-d}&0&0\\0&0&\mathds{1}_{k-(10-d)}\end{pmatrix}.
\end{align}
Note that the $10-d$ graviphotons are excluded from $k$. Let us take the gauge group to be $G=U(1)^{r-1}\times SU(2)$, and move in its Coulomb branch to Higgs it into $U(1)^r$. This allows us to see how the two massive vector multiplets $W_\pm$ can supersymmetrically couple to the other massless multiplets. In this example, $k=r+2$. The bosonic part of the action is given by \cite{Hohm:2014sxa}

\begin{align}
    S=\int d^nx \sqrt{-g} e^{-2\phi}[\mathcal{R}(g)&+4\partial_\mu\phi\partial^\mu\phi-\frac{1}{12}H^{\mu\nu\rho}H_{\mu\nu\rho}\nonumber\\
    &+\frac{1}{8}D^\mu \hat{\mathcal{H}}^{\hat M \hat N}D_\mu \hat{\mathcal{H}}_{\hat M \hat N}-\frac{1}{4}\hat{\mathcal{H}}_{\hat M \hat N}\hat{\mathcal{F}}^{\mu\nu\hat{M}}\tensor
    {\hat{\mathcal{F}}}{_\mu_\nu^{\hat{N}}}-V(\mathcal{H})],
\end{align}
where
\begin{align}
    D_\mu \hat{\mathcal{H}}^{\hat M \hat N}&=\partial_\mu\hat{H}^{\hat{M} \hat{N}}-2\tensor{\hat{A}}{_\mu^{\hat{K}}}\tensor{f}{_{\hat K}^{(\hat M}_{\hat L}}\tensor{\hat{\mathcal{H}}}{^{\hat N)}^{\hat L}}.\,\nonumber\\
    \tensor{\hat{\mathcal{F}}}{_{\mu\nu}^{\hat{M}}}&=2\partial_{[\mu}\tensor{\hat{A}}{_{\nu]}^{\hat{M}}}+\tensor{f}{^{\hat M}_{\hat K}_{\hat L}}\tensor{\hat{A}}{_{\mu}^{\hat{K}}}\tensor{\hat{A}}{_{\nu}^{\hat{L}}},\nonumber\\
    H_{\mu\nu\rho}&=3(\partial_{[\mu]}B_{\nu\rho]}-\tensor{\hat{A}}{_{[\mu}^{\hat M}}\partial_\nu\tensor{\hat{A}}{_{\rho]}_{\hat M}} -\frac{1}{3}\tensor{f}{_{\hat M}_{\hat K}_{\hat L}}\tensor{\hat{A}}{_{[\mu}^{\hat{M}}}\tensor{\hat{A}}{_{\nu}^{\hat{K}}}\tensor{\hat{A}}{_{\rho]}^{\hat{L}}}),\nonumber\\
    V(\mathcal{H})&=\tensor{f}{^{\hat M}_{\hat K}_{\hat P}}\tensor{f}{^{\hat N}_{\hat L}_{\hat Q}}\hat{\mathcal{H}}_{\hat M \hat N}\hat{\mathcal{H}}^{\hat K \hat L}\hat{\mathcal{H}}^{\hat P \hat Q}+\frac{1}{4}\tensor{f}{^{\hat M}_{\hat N}_{\hat K}}\hat{\mathcal{H}}_{\hat M \hat N}\tensor{f}{^{\hat N}_{\hat M}_{\hat L}}\hat{\mathcal{H}}^{\hat K \hat L}+\frac{1}{6}\tensor{f}{_{\hat M}_{\hat N}_{\hat K}}\tensor{f}{^{\hat M}^{\hat N}^{\hat K}},
\end{align}
where the scalars $\hat{\mathcal{H}}^{\hat K \hat L}$ are in $O(10-d,r+2)/[O(10-d)\times O(r+2)]$, and $\hat{M}=1,\cdots,10-d+r+2$.

Suppose that $SU(2)$ corresponds to indices $10-d+r\leq\hat M\leq10-d+r+2$. After Higgsing the $SU(2)$, we can use the quotient group $O(10-d)\times O(r+2)$ to make all the scalars in $\ln[\hat{\mathcal{H}}]^{\hat M \hat N}$ where $\max\{\hat M,\hat N\}\geq 10-d+r$ except the following vanish.
\begin{align}
    (\hat M,\hat N)\in&\{1,\hdots2(10-d)\}\times\{10-d+r,10-d+r+1,10-d+r+2\}\nonumber\\
   &\cup\{10-d+r,10-d+r+1,10-d+r+2\}\times\{1,\hdots2(10-d)\}.
\end{align}
such that for $10-d<\hat M\leq 2(10-d)$ and $\hat N\in \{10-d+r,10-d+r+1,10-d+r+2\}$ we have 
\begin{align}
    \ln[\hat{\mathcal{H}}^{\hat M \hat N}]=-\ln[\hat{\mathcal{H}}^{\hat M-(10-d) ~~\hat N}]=\ln[\hat{\mathcal{H}}^{\hat N \hat M}]=-\ln[\hat{\mathcal{H}}^{\hat N ~~\hat M-(10-d)}].
\end{align}
We use $\ln(\hat{\mathcal{H}})$ because it belongs to the Lie algebra of $O(d)\times O(r+2)$, which has a simpler description. 

We can also use the gauge symmetry to impose the unitary gauge $\ln[\hat{\mathcal{H}}]^{\hat N ~11-d}=\ln[\hat{\mathcal{H}}]^{11-d~ \hat N}=0$ for $\hat N> 10-d+r$ and $\ln[\hat{\mathcal{H}}]^{10-d+r ~11-d}=C$ is the Higgsing parameter that sets the mass scale of the Higgsed vector multiplets $\hat M_\text{Massive}\in\{10-d+r+1,10-d+r+2\}$. Note that these massive vector multiplets now each have $9-d$ scalars corresponding to $\ln[\hat{\mathcal{H}}]^{\hat M \hat N}$ where $11-d<\hat N\leq2(10-d)$.

Looking at the action, one can see that the charge of the two Higgsed vector multiplet ${\hat A}^{\hat N}$ under the massless vector ${\hat A}^{\hat M}$ comes from the terms quadratic in $\mathcal{F}$ and is proportional to ${\hat{\mathcal{H}}}^{\hat M \hat N}$. This scalar depends on scalars in the massless and massive multiplets through exponentiation of $\ln[\hat{\mathcal{H}}]$. Even though the dependence of the gauge couplings on the scalars in the massless multiplets is complicated, its change under the change of them is easy. Changing the scalars in the massless multiplets corresponds to a similarity transformation on $\hat{\mathcal{H}}$ by an element of $O(10-d,r)$, which acts on the first $10-d+r$ indices. Therefore, the charges of the massive multiplets which are given by $\hat{\mathcal{H}}^{\hat M \hat N}$ with $\hat{M}\leq 10-d+r<\hat{N}$ transform in the fundamental representation of $O(10-d,r)$. 
\begin{align}
    \Lambda\in O(10-d,r): \hat{\mathcal{H}}^{\hat M \hat N}\rightarrow \Lambda^{\hat {M}'}_{\hat{M}}\hat{\mathcal{H}}^{{\hat M}' \hat N}.
\end{align}
In fact, this argument holds for any Higgsed massive multiplet. Consequently, if the theory has a point of maximal enhancement where the gauge algebra becomes simple, this argument tells us that the charge lattice transforms covariantly under $O(10-d,r)$ changes of the moduli. We are particularly interested in 9d supergravities, which according to Swampland arguments \cite{Bedroya:2021fbu}, always have such a point of symmetry enhancement. However, even if such a point did not exist, such a constraint is generally expected from supersymmetry. For the coupling terms ${\hat A}^{\hat M}\times\hdots$ or $\partial {\hat A}^{\hat M}\times\hdots$ to be supersymmetrically invariant, we need to cancel the second order variation of ${\hat A}^{\hat M}$ under supersymmetry with terms which involve scalars in the same multiplet as ${\hat A}^{\hat M}$, which is why we typically end up with a term like $\mathcal{H}_{\hat M \hat N}{\hat A}^{\hat M}\times\hdots$ or $\mathcal{H}_{\hat M \hat N}\partial{\hat A}^{\hat M}\times\hdots$. Even though the rest of the terms (such as the mass term) can have different scalar dependencies, the coupling to the massless vector multiplets is expected to have fixed dependence on the scalars in the massless multiplets due to supersymmetry.

\bibliographystyle{TitleAndArxiv}
\bibliography{References}
\end{document}